\documentstyle [12pt] {article}
\begin{document}
\def\be{\begin{eqnarray}}
\def\en{\end{eqnarray}}
\def\non{\nonumber}
\def\la{\langle}
\def\ra{\rangle}
\def\ep{\varepsilon}
\def\b{\Lambda_b\to J/\psi\Lambda}
\def\j{{J/\psi}}
\def\Ri{\Rightarrow}
\def\bb{{\bar{B}}}
\def\pr{{\sl Phys. Rev.}~}
\def\prl{{\sl Phys. Rev. Lett.}~}
\def\pl{{\sl Phys. Lett.}~}
\def\np{{\sl Nucl. Phys.}~}
\def\zp{{\sl Z. Phys.}~}

\font\el=cmbx10 scaled \magstep2
{\obeylines
\hfill IP-ASTP-23-94
\hfill November, 1994}

\vskip 1.5 cm

\centerline{\large\bf Vector Dominance Effects in Weak Radiative Decays}
\centerline{\large\bf of the $B$ Mesons}
\medskip
\bigskip
\medskip
\centerline{\bf Hai-Yang Cheng}
\medskip
\centerline{ Institute of Physics, Academia Sinica}
\centerline{Taipei, Taiwan 11529, Republic of China}
\bigskip
\bigskip
\bigskip
\centerline{\bf Abstract}
\bigskip
{\small  The long distance vector-meson-dominance (VMD) effects on the weak
radiative decays $\bb\to\rho\gamma$ and $\bb^0\to D^{*0}\gamma$ are studied.
For $\bb\to\rho\gamma$ decays, the VMD contribution is $(10-20)\%$ of
the short-distance penguin amplitude. The pole effect is as important as the
VMD one in the decay $B^-\to\rho^-\gamma$, but it is suppressed in $\bb^0\to
\rho^0\gamma$. The branching ratio of $\bb\to\rho\gamma$, estimated to be of
order $10^{-6}$, strongly depends on the sign of the Wolfenstein parameter
$\rho$. A measurement of any deviation of the ratio $R=\Gamma(B^-\to\rho^-
\gamma)/\Gamma(\bb^0\to\rho^0\gamma)$ away from the isospin value 2 will not
only provide a probe on the long-range contribution but also fix
the sign of $\rho$: $R>2$ for $\rho<0$ and $R<2$ for $\rho>0$.
The decay $\bb^0\to D^{*0}\gamma$ does not receive short-distance
contributions, and its branching ratio, predicted to be $0.9\times 10^{-6}$,
is dominated by $W$-exchange accompanied by a photon emission.
}

\pagebreak
\noindent {\bf 1. Introduction}

Recently the weak radiative decays of $B$ mesons and bottom baryons have been
systematically
studied in Ref.[1]. At the quark level, there are two essential mechanisms
responsible for weak radiative decays: electromagnetic penguin mechanism
and $W$-exchange (or $W$-annihilation) bremsstrahlung. The two-body decays of
the $B$ meson
proceeding through the short-distance electromagnetic penguin diagrams are:
\be
&& b\to s\gamma\Ri~\bar{B}\to\bar{K}^*\gamma,~~\bar{B}_s\to\phi\gamma, \non \\
&&b\to d\gamma\Ri~\bar{B}\to\rho\gamma,~~\bb^0\to\omega\gamma,~~\bar{B}_s\to
K^{*0}\gamma,
\en
while the decay modes occurring through $W$-exchange or $W$-annihilation
accompanied by a photon emission are:
\be
&&W{\rm -exchange}:\cases{b\bar{d}\to c\bar{u}\gamma\Ri~\bar{B}^0\to D^{*0}
\gamma, & \cr b\bar{s}\to c\bar{u}\gamma\Ri~\bar{B}_s\to D^{*0}\gamma,~~~~~b
\bar{d}\to c\bar{c}\gamma\Ri~\bar{B}^0\to\j\gamma, \cr} \non \\
&& W{\rm -annihilation}:~~b\bar{u}\to s\bar{c}\gamma\Ri~B^-\to D_s^{*-}\gamma,
{}~~~~~b\bar{u}\to d\bar{c}\gamma\Ri~B^-\to D^{*-}\gamma.
\en
Note that decay modes in (1) also receive contributions from $W$-exchange
or $W$-annihilation bremsstrahlung, but they are in general quark mixing
suppressed.

    At the hadronic level, the $W$-exchange diagrams manifest as long-distance
pole diagrams. However, another possible long-distance effect, namely
vector meson dominance (VMD) contribution, was advocated sometime ago by
Golowich and Pakvasa [2]. For example, $B\to K^*\gamma$ can proceed through
$B\to K^*\j\to K^*\gamma$ via $\j-\gamma$ conversion. Since the concept of VMD
though useful has never been derived from the standard model, it is not clear
at all if this VMD contribution to $B\to V\gamma$ is a real one. In fact, it
has been argued that at the quark level $b\to s\j\to s\gamma$ is not allowed
at the tree level because of gauge invariance [3]. It is also easily seen at
the hadronic level that for a given $B\to VV'$ amplitude with $V'$ being a
neutral vector meson, it is no longer gauge invariant after a replacement
of the polarization vector $\ep_\mu(V')$ of the vector meson $V'$ by the photon
one $\ep_\mu(\gamma)$. This is ascribed to the fact that, as elaborated on in
Refs.[4,5], the helicity amplitude of $B\to VV'$ has a longitudinal component
that spoils gauge invariance after $V'-\gamma$ conversion. Therefore, in
order to retain gauge invariance, one must disregard the longitudinal
helicity amplitude of $B\to VV'$ for a correct usage of VMD [4,5].

   In the present paper we will assume the validity of VMD and estimate its
effect on weak radiative decays. To be specific, we will consider two
representative decay modes in (1) and (2): $\bar{B}\to\rho\gamma$ and
$\bar{B}^0\to D^{*0}\gamma$. A generalization of the present work to other
radiative decays is straightforward.
\vskip 0.4cm

\noindent  {\bf 2. The $\bb\to\rho\gamma$ Decay}

The radiative decay $\bb\to\rho\gamma$
is of experimental and theoretical intertest since we may learn the quark
mixing matrix element $V_{td}$ from its measurement [6]. This decay resembles
to $B\to K^*\gamma$ in many ways. It is well known that the latter
is dominated
by the short-distance electromagnetic penguin $b\to s\gamma$. There are two
possible long-distance effects: VMD and $W$-exchange bremsstrahlung; the latter
manifested as a long-distance pole contribution at the hadronic level. A recent
estimate gives [4]
\be
\left|{A_{\rm VMD}\over A_{\rm expt}}\right|_{B\to K^*\gamma}\leq 0.1\,,~~~~~
\left|{A_{\rm pole}\over A_{\rm expt}}\right|_{B\to K^*\gamma}\simeq 0.01\,.
\en
The long-distance contribution is thus dominated by the VMD effect, arising
mainly from the process $B\to K^*\j\to K^*\gamma$. The pole contribution
is suppressed due to the smallness of the
weak mixing $V_{ub}V_{us}$.  Apart from the mixing angles, the decay $\bb\to
\rho\gamma$ proceeds in the same way as $B\to K^*\gamma$. In this section,
we will estimate the short- and long-distance contributions to $\bb\to\rho
\gamma$ and see if the pattern (3) is still respected. An estimate of the
long-distance effect on $\bb\to\rho\gamma$ was recently made in Ref.[7].
We will present in this paper a more quantitative study.

The general amplitude of weak
radiative decay with one real photon emission is given by
\be
A[\bar{B}(p)\to P^*(q)\gamma(k)] & = & i\epsilon_{\mu\nu\alpha\beta}\ep
^\mu k^\nu\ep^{*\alpha}q^\beta f_1(k^2)  \non \\
& + & \ep^\mu[\ep^*_\mu(m^2_B-m^2_{P^*})-(p+q)_\mu\ep^*\cdot k]
f_2(k^2),
\en
where $\ep$ and $\ep^*$ are the polarization vectors of the photon
and the vector meson $P^*$, respectively, the first (second) term on the r.h.s.
is parity conserving (violating), and $k^2=0$. The decay width implied by
the amplitude (4) is
\be
\Gamma(\bar{B}\to P^*\gamma)=\,{1\over 32\pi}\,{(m^2_B-m^2_{P^*})^3\over
m^3_B}\,(|f_1(0)|^2+4|f_2(0)|^2).
\en

   To begin with, we consider the transition amplitude induced by the
short-distance penguin $b\to d\gamma$
\be
A(b\to d\gamma)=\,i{G_F\over\sqrt{2}}\,{e\over 8\pi^2}\left(\sum_i F_2(x_i)V_
{ib}V_{id}^*\right)
\ep^\mu k^\nu\bar{d}\sigma_{\mu\nu}[m_b(1+\gamma_5)+m_d(1-\gamma_5)]b,
\en
where $x_i=m_i^2/M_W^2$, $m_i$ is the mass of the quark $i$, and $F_2$ is
a smooth function of $x_i$ [8]. In the static limit of the
heavy $b$ quark, we may use the equation of motion $\gamma_0 b=b$ to derive
the relation [9]
\be
\la\rho|\bar{d}i\sigma_{0i}(1\pm\gamma_5)b|B\ra =\,\la \rho|
\bar{d}\gamma_i(1\mp\gamma_5)b|B\ra.
\en
As a result, the form factors $f_1$ and $f_2$ in (4)
can be related to the vector and axial-vector form factors $V$ and $A_1$
appearing in the matrix element on the r.h.s. of Eq.(7) defined by [10]
\be
\la\rho(p_\rho)|\bar{d}\gamma_\mu b|B(p_B)\ra &=& {2i\over m_B+m_\rho}\epsilon
_{\mu\nu\alpha\beta}\ep^{*\nu}p_\rho^\alpha p^\beta_B V^{B\rho}(q^2),  \non \\
\la\rho(p_\rho)|\bar{d}\gamma_\mu\gamma_5b|B(p_B)\ra &=& (m_B+m_\rho)\ep^*_\mu
A_1^{B\rho}
(q^2)-{\ep^*\cdot q\over m_B+m_\rho}(p_B+p_\rho)_\mu A_2^{B\rho}(q^2)  \\
&&-{2\ep^*\cdot q\over q^2}q_\mu m_\rho[A_3^{B\rho}(q^2)-A_0^{B\rho}(q^2)],
\non
\en
with $q=p_B-p_\rho$. At $k^2=0$, we obtain (see e.g. Ref.[11])
\be
f_1^{\rm peng}(B^-\to\rho^-\gamma) &=& -{G_F\over\sqrt{2}}\,{e\over 8\pi^2}
\left(\sum_i F_2(x_i)V_{ib}V^*_{id}\right)m_bF^{B\rho}(0)  \non \\
f_2^{\rm peng}(B^-\to\rho^-\gamma) &=& -{1\over 2}f_1^{\rm peng}(B^-\to\rho^-
\gamma),
\en
and
\be
f_i^{\rm peng}(\bb^0\to\rho^0\gamma)=-{1\over\sqrt{2}}f_i^{\rm peng}(B^-\to
\rho^-\gamma),
\en
where
\be
F^{B\rho}(0)=\, {m_B-m_\rho\over m_B}V^{B
\rho}(0)+{m_B+m_\rho\over m_B}A_1^{B\rho}(0).
\en
Two remarks are in order. (i) Eq.(9) is subject to ${\cal O}(1/m_b)$
corrections which are not included here. (ii) Apart from the quark mixing
angles, the short-distance $\bb\to\rho\gamma$ amplitude is different from
the $B\to K^*\gamma$ one in that the $u$ quark loop contribution is
negligible in the latter but not necessarily so in the former.
To be precise, the $B\to K^*\gamma$ amplitude is given by
\be
f_1^{\rm peng}(B\to K^*\gamma) &\cong& -{G_F\over\sqrt{2}}\,{e\over 8\pi^2}\,
F_2(x_t)V_{tb}V^*_{ts}m_bF^{BK^*}(0), \non \\
f_2^{\rm peng}(B\to K^*\gamma) &=& -{1\over 2}f_1^{\rm peng}(B\to K^*\gamma),
\en
where uses of the approximations $F_2(x_t)-F_2(x_c)\cong F_2(x_t)$ and
$V_{cb}V^*_{cs}\approx -V_{tb}V^*_{ts}$ due to
the smallness of $V_{ub}V_{us}^*$
have been made. Numerically, $F_2(x_t)=0.65$
for $\Lambda_{\rm QCD}=200$ MeV and $m_t=174$ GeV.
It follows from (9)-(12) that the short-distance $\bb\to\rho\gamma$ and
$B\to K^*\gamma$ amplitudes are related by
\be
f_i^{\rm peng}(B^-\to\rho^-\gamma)=\,{V_{td}^*\over V_{ts}^*}(1+\Delta)\,{F^{
B\rho}(0)\over F^{BK^*}(0)}f_i^{\rm peng}(B\to K^*\gamma),
\en
with
\be
\Delta=\,{F_2(x_u)-F_2(x_c)\over F_2(x_t)-F_2(x_c)}\,{V_{ub}\over V_{td}^*}.
\en
For later purposes of numerical estimate, we will follow Ref.[7] to take
$(F_2(x_u)-F_2(x_c))/(F_2(x_t)-F_2(x_c))\simeq -0.30$.

   We next turn to long-distance contributions and first focus on the VMD part.
The transitions $\bb\to \rho V$ followed by $V-\gamma$ conversion are
dominated by the virtual vector mesons $V=\j,~\psi',~\rho^0$ and $\omega$
as depicted in Figs.1 and 2.
To illustrate the use of VMD, let us consider the hadronic decay $B^-\to\rho^-
\j$ as an example. Assuming factorization, its amplitude reads
\be
A(B^-\to \rho^-\j)=\,{G_F\over\sqrt{2}}V_{cb}V_{cd}^*a_2\ep^{\mu}(\j)
\ep^{*\nu}(\rho)\left(\hat{A}_1g_{\mu\nu}+\hat{A}_2p_\mu^B p_\nu^B+i\hat{V}
\epsilon_{\mu\nu\alpha\beta}p^\alpha_\rho p^\beta_{B}\right),
\en
where
\be
\hat{A}_1 &=& -(m_B+m_\rho)f_\j m_\j A_1^{B\rho}(m^2_\j),  \non \\
\hat{A}_2 &=& {2\over m_B+m_\rho}f_{\j}m_\j A_2^{B\rho}(m^2_{\j}), \\
\hat{V} &=& {2\over m_B+m_\rho}f_{\j}m_{\j}V^{B\rho}(m^2_{\j}), \non
\en
and $a_2$ is a parameter introduced in Ref.[12] for the internal $W$-emission
diagram. VMD implies that a possible contribution to $B^-\to\rho^-\gamma$
comes from the decay $B^-\to\rho^-\j$ followed by continuing its amplitude
from $p_\j^2=m^2_\j$ to $p^2_\j=0$ and replacing the vector-meson's
polarization vector $\ep_\mu(\j)$ by the photon one:
\be
\ep_\mu(V)\to {e\over g_{\gamma V}}\ep_\mu(\gamma),
\en
where $g_{\gamma V}$ is a dimensionless quantity defined by
\be
\la 0|J_\mu^{\rm em}|V\ra=\,{m^2_V\over g_{\gamma V}}\ep_\mu.
\en
In order to retain gauge invariance of the $B^-\to\rho^-\gamma$ amplitude,
it becomes necessary to demand a vanishing $A(B^-\to\rho^-\gamma)_{\rm VMD}$
when $\ep_\mu(\gamma)\to k_\mu$. This is equivalent to discarding
the longitudinal polarization component of
the $B^-\to\rho^-\j$ amplitude in the $p^2_\j\to 0$ limit [4,5];
\footnote{For the process such as $B\to K^*\j\to K^*\gamma$, one may employ
the experimental measurement of the transverse polarization component of
$B\to K^*\j$ to compute the VMD contribution to $B\to K^*\gamma$. In the
absence of experimental information for $\bb\to\rho\j$ etc., we have to appeal
to some model calculations for evaluating the VMD $\bb\to\rho\gamma$ amplitude,
as we have done here.}
that is,
\be
\hat{A}_1+(p_B\cdot k)\hat{A}_2=\,\hat{A}_1+{1\over 2}(m_B^2-m_\rho^2)\hat{A}
_2=0.
\en
Substituting (19) into (15) yields
\be
A(B^-\to\rho^-\j\to\rho^-\gamma) &=& {e\over g_{\gamma\j}}\,{G_F\over\sqrt{2}}
V_{cb}V_{cd}^*\Big\{i\epsilon_{\mu\nu\alpha\beta}\ep^\mu k^\nu \ep^
{*\alpha}p_\rho^\beta\hat{V}   \non \\
&&-{1\over 2}\ep^\mu[\ep^*_\mu(m_B^2-m^2_\rho)-(p_B+p_\rho)_\mu \ep^*\cdot k]
\hat{A}_2\Big\}.
\en
Comparing (20) with (4), assuming SU(3)-flavor symmetry for heavy-light form
factors and summing over the intermediate vector meson
states gives rise to (see Fig.1)
\be
f_1^{\rm VMD}(B^-\to\rho^-\gamma) &=& eG_F\Bigg\{\sqrt{2}V_{cb}V_{cd}^*a_2{
1\over m_B+m_\rho}\left({f_\j m_\j\over g_{\gamma\j}}+{f_{\psi'}m_{\psi'}\over
g_{\gamma\psi'}}\right)  \non \\
&& +V_{ub}V_{ud}^*a_1f_\rho m_\rho\left({1\over m_B+m_\rho}{1\over g_{\gamma
\rho}}+{1\over m_B+m_\omega}{1\over g_{\gamma\omega}}\right)\Bigg\}V^
{B\rho}(0),  \\
f_2^{\rm VMD}(B^-\to\rho^-\gamma) &=& -{1\over 2}{A_2^{B\rho}(0)\over
V^{B\rho}(0)}f_1^{\rm VMD}(B^-\to\rho^-\gamma),   \non
\en
where $a_1$ is a parameter introduced for the external $W$-emission diagram
[12], and the relative sign between $\rho^0$- and $\omega$-mediated VMD
amplitudes is fixed by the wave functions $\rho^0={1\over\sqrt{2}}(\bar{u}u
-\bar{d}d)$ and $\omega={1\over\sqrt{2}}(\bar{u}u+\bar{d}d)$.
Likewise, for $\bb^0\to\rho^0\gamma$ decay (see Fig.2):
\be
f_1^{\rm VMD}(\bb^0\to\rho^0\gamma) &=& -{1\over\sqrt{2}}eG_F\Bigg\{\sqrt{2}V_
{cb}V_{cd}^*a_2{
1\over m_B+m_\rho}\left({f_\j m_\j\over g_{\gamma\j}}+{f_{\psi'}m_{\psi'}\over
g_{\gamma\psi'}}\right)  \non \\
&& +V_{ub}V_{ud}^*a_2f_\rho m_\rho\left({1\over m_B+m_\rho}{1\over g_{\gamma
\rho}}-{1\over m_B+m_\omega}{1\over g_{\gamma\omega}}\right)\Bigg\}V^{B\rho}
(0),   \\
f_2^{\rm VMD}(\bb^0\to\rho^0\gamma) &=& -{1\over 2}{A_2^{B\rho}(0)\over
V^{B\rho}(0)}f_1^{\rm VMD}(\bb^0\to\rho^0\gamma). \non
\en
Note that the isospin relation for the decay rates
$\Gamma(\bb^0\to\rho^0\gamma)={1\over {2}}\Gamma(B^-
\to\rho^-\gamma)$, respected by the short-distance penguin interaction
[see Eq.(10)], is no longer satisfied by the VMD contributions arising from
$\rho^0$ and $\omega$ intermediate states as the decays $\bb^0\to\rho^0
\rho^0,~\rho^0\omega$ are color suppressed, while $B^-\to\rho^-\rho^0,~\rho^-
\omega$ are not.

   We now come back to the coupling $g_{\gamma V}$ defined in Eq.(18). In the
quark model, $g_{\gamma V}$ is proportional to $\sum_i a_ie_i$ with $a_i$
being the coefficient of the $i$th quark with charge $e_i$ in the wave
function. Consequently, it is expected that
\be
g_{\gamma\rho}^{-1}:g_{\gamma\omega}^{-1}:g_{\gamma\phi}^{-1}=\,3:1:-\sqrt{2}.
\en
Experimentally, $g_{\gamma V}$ can be determined from the measured $V\to\ell^+
\ell^-$ rate:
\be
\Gamma(V\to\ell^+\ell^-)=\,{4\pi\alpha^2\over 3}\,{m_V\over g_{\gamma V}^2}
\left(1-4{m_\ell^2\over m_V^2}\right)^{1/2}\left(1+2{m_\ell^2\over m_V^2}
\right).
\en
{}From the measured widths [13] we obtain
\be
g_{\gamma\rho}=5.05,~~g_{\gamma\omega}=17.02,~~g_{\gamma\phi}=-12.89,~~
g_{\gamma\j}=11.75,~~g_{\gamma\psi'}=18.87,
\en
where we have applied the quark model to fix the sign. Therefore, the
relation (23) is satisfied experimentally. The vector-meson decay constant
$f_V$ is related to $g_{\gamma V}$ via the relation
\be
f_V=\,m_V(g_{\gamma V}\sum_i a_ie_i)^{-1}.
\en
It follows that the decay constants relevant to our purposes are
\footnote{To determine $f_\j$ and $f_{\psi'}$ we have taken into account
the momentum dependence of the fine structure constant.}
\be
f_\rho=216\,{\rm MeV},~~~f_\j=395\,{\rm MeV},~~~f_{\psi'}=293\,{\rm MeV}.
\en

    Another long-distance contribution to $\bb\to\rho\gamma$ stems from
the $W$-annihilation diagram for $B^-\to\rho^-\gamma$ and the $W$-exchange
diagram for $\bb^0\to\rho^0\gamma$ (see Fig.3).
\footnote{Contrary to Ref.[7], we count the VMD and pole effects as
two different long-range contributions
(see also Ref.[4]). As we shall see from Table I,
while VMD and pole contributions to $B^-\to \rho^-\gamma$ are comparable,
they are different by one order of magnitude in the $\bb^0\to\rho^0\gamma$
decay amplitude.}
Using the formulism developed in Sec. II of Ref.[1],
the pole contributions are found to be
\footnote{Strictly speaking, the formulism developed in Sec.II of Ref.[1]
is applicable only if both initial and final hadrons can be treated as heavy.
Nevertheless, we believe that (28) and (29) here are good for
order-of-magnitude estimate. Basically, our approach is similar to the second
method advocated in Ref.[7]. As stressed in the Introduction, $W$-exchange
(or $W$-annihilation) contributions manifest as pole diagrams at
the hadronic level. This equivalence has been demonstrated explicitly for
$\bb^0\to D^{*0}\gamma$ in Ref.[1].}
\be
f_1^{\rm pole}(B^-\to\rho^-\gamma) &=& \kappa a_1\left[\left({e_d\over m_d}+
{e_u\over m_u}\right){m_\rho\over m_B}+\left({e_u\over m_u}+{e_b\over m_b}
\right)\right]{m_B m_\rho\over m_B^2-m^2_\rho},  \non \\
f_2^{\rm pole}(B^-\to\rho^-\gamma) &=& -{1\over 2}\kappa a_1\left[\left({e_d
\over m_d}-{e_u\over m_u}\right){m_\rho\over m_B}+\left({e_u\over m_u}-
{e_b\over m_b}\right)\right]{m_B m_\rho\over m_B^2-m^2_\rho},
\en
and
\be
f_1^{\rm pole}(\bb^0\to\rho^0\gamma) &=& {1\over\sqrt{2}}\kappa a_2\left[
2{e_u\over m_u}{m_\rho\over m_B}+\left({e_d\over m_d}+{e_b\over m_b}
\right)\right]{m_B m_\rho\over m_B^2-m^2_\rho},  \non \\
f_2^{\rm pole}(\bb^0\to\rho^0\gamma) &=& -{1\over 2\sqrt{2}}\kappa a_2
\left({e_d\over m_d}-{e_b\over m_b}\right){m_B m_\rho\over m_B^2-m^2_\rho},
\en
where $\kappa=eG_FV_{ub}V_{ud}^*f_B f_\rho/\sqrt{2}$,
 and $m_i$ is the constituent quark mass. Again, we see that isospin
symmetry is violated as the $W$-exchange amplitude is color suppressed whereas
$W$-annihilation is color favored.

\vskip 0.4 cm
\noindent   {\bf 3. The $\bb^0\to D^{*0}\gamma$ Decay}

The radiative decay $\bb^0\to D^{*0}\gamma$ receives
only long-distance contributions, and yet its branching ratio is large
enough for a feasible test in the near future. In Ref.[1] an effective
Lagrangian for the quark-quark bremsstrahlung $b\bar{d}\to c\bar{u}\gamma$
is derived based on the fact that the intermediate quark state in this process
is sufficiently off-shell and the emitted photon is hard enough, allowing
an analysis of the $W$-exchange bremsstrahlung by perturbative QCD. Applying
this formulism to $\bb^0\to D^{*0}\gamma$ yields (see (3.7) of Ref.[1])
\be
f_1^{\rm pole}(\bb^0\to D^{*0}\gamma) &=& \kappa'a_2\left[ \left({e_c\over
m_c}+{e_u\over m_u}\right){m_{D^*}\over
m_B}+\left({e_d\over m_d}+{e_b\over m_b}\right)\right]\,{m_Bm_{D^*}\over
m_B^2-m_{D^*}^2}, \non \\
f_2^{\rm pole}(\bb^0\to D^{*0}\gamma) &=& -{1\over 2}\kappa' a_2\left[ \left(
{e_c\over m_c}-{e_u\over m_u}\right){m_{D^*}\over
m_B}+\left({e_d\over m_d}-{e_b\over m_b}\right)\right]\,{m_Bm_{D^*}\over
m_B^2-m_{D^*}^2},
\en
with $\kappa'=\,eG_FV_{cb}V_{ud}^*f_Bf_{D^*}/\sqrt{2}$.
It has been shown
explicitly in Ref.[1] that the effective Lagrangian and pole model
approaches are equivalent, but the former is much simpler and provides
information on the form factors.

   It is easily seen that the VMD contributions to $\bb^0\to D^{*0}\gamma$
come from the processes $\bb^0\to D^{*0}\rho^0(\omega)\to D^{*0}\gamma$.
Following Sec. II, we obtain
\be
f_1^{\rm VMD}(\bb^0\to D^{*0}\gamma) &=& -eG_FV_{cb}V_{ud}^*a_2f_{D^*}m_{D^*}
\left({1\over m_B+m_\rho}{1\over g_{\gamma\rho}}-{1\over m_B+m_\omega}{1\over
g_{\gamma\omega}}\right)V^{B\rho}(0),  \non \\
f_2^{\rm VMD}(\bb^0\to D^{*0}\gamma) &=& -{1\over 2}{A_2^{B\rho}(0)\over
V^{B\rho}(0)}f_1^{\rm VMD}(\bb^0\to D^{*0}\gamma).
\en
Numerical results will be presented in the next section.

\vskip 0.4 cm
\noindent  {\bf 4. Numerical Results}

   To estimate the short-distance penguin, long-distance VMD and pole
contributions to weak radiative decays, we will use the following values for
various quantities:

 (i) decay constants for pseudoscalar and vector mesons. In addition to
Eq.(27), we also use
\be
f_B=190\,{\rm MeV},~~~f_{D^*}=200\,{\rm MeV}.
\en

(ii) $a_1$ and $a_2$. The parameters $a_1$ and $a_2$ appearing in nonleptonic
$B$ decays are recently extracted from the CLEO data [14] of $B\to
D^{(*)}\pi(\rho)$ and $B\to \j K^{(*)}$ to be [15]
\be
&&a_1(B\to D^{(*)}\pi(\rho)) = 1.01\pm 0.06,~~~a_2(B\to D^{(*)}\pi(\rho))=0.23
\pm 0.06, \non \\
&& \left|a_2(B\to\j K^{(*)})\right| = 0.227\pm 0.013.
\en
Hence, in the present paper it is natural to employ
\be
a_1=1.01,~~~~~a_2=0.23\,.
\en

(iii) photon-vector meson coupling constants given by Eq.(25).

(iv) constituent quark masses:
\be
m_u=338\,{\rm MeV},~~~m_d=322\,{\rm MeV},~~~m_c=1.6\,{\rm GeV},~~~m_b=5\,
{\rm GeV},
\en
where the light quark masses are taken from p. 1729 of Ref.[13].

(v) form factors $A_{1,2}$ and $V$ at $q^2=0$. It has been shown in Ref.[15]
that the heavy-flavor-symmetry approach for heavy-light form
factors in
conjunction with a certain type of form-factor $q^2$ dependence provides
a satisfactory description of the CLEO data for the ratio $\Gamma(B\to\j K^*)/
\Gamma(B\to \j K)$ [14] and the CDF measurement of the fraction of longitudinal
polarization in $B\to \j K^*$ [16]. Assuming SU(3)-flavor
symmetry for heavy-light form factors, we find from Table I of Ref.[15] that
\be
V^{B\rho}(0)=0.33,~~~A_1^{B\rho}(0)=0.29,~~~A_2^{B\rho}(0)=0.19.
\en
As shown in Ref.[15], the CLEO measurement ${\cal B}(B\to K^*\gamma)=(4.5\pm
1.5\pm 0.9)\times 10^{-5}$ [17] is well explained by the same set of form
factors.

(vi) quark mixing matrix elements. We will take $V_{cb}=0.040$ [18] and
$|V_{ub}/V_{cb}|=0.08$, which in turn imply the following Wolfenstein
parameters ($\lambda=0.22$) [19]:
\be
A=0.826,~~~~~~\sqrt{\rho^2+\eta^2}=0.35.
\en
For the purpose of illustration, we will take $\eta=0.30$, and hence $\rho=\pm
0.18$. A small and negative $\rho$ is favored by $B^0-\bar{B}^0$ mixing data.
In terms of the Wolfenstein parametrization of the quark mixing matrix [19],
the quantity appearing in Eq.(13) has the expression
\be
{V_{td}^*\over V_{ts}^*}(1+\Delta)\cong -\lambda(1-1.3\rho+i1.3\eta).
\en

   With the values given by (33)-(38) for various quantities, we proceed to
compute the form factors $f_1$ and $f_2$ for $\bb\to\rho\gamma$ and
$\bb^0\to D^{*0}\gamma$ decays; their explicit
expressions are shown in Secs. II and III. It should be stressed that
all the relative signs among various amplitudes are fixed in our work.
The ratios of the long-distance (VMD and pole) and short-distance
(penguin) contributions to $f_{1,2}$ in $\bb\to\rho\gamma$ are summarized
in Table I. We see from Table I that
while VMD and pole amplitudes are comparable in $B^-\to\rho^-\gamma$ decay,
estimated to be roughly $(10-20)\%$ of the short-distance
contribution, the former is the dominant long-distance contribution to
$\bb^0\to\rho^0\gamma$. Taking into accout various contributions to $f_{1,2}$
\be
f_{1,2}^{\rm tot}=\,f_{1,2}^{\rm peng}+f_{1,2}^{\rm VMD}+f_{1,2}^{\rm pole},
\en
we obtain the branching ratios
\be
{\cal B}(B^-\to\rho^-\gamma)=\cases{1.5\times 10^{-6},  \cr 4.4\times 10^{-6},
\cr}~~~~~{\cal B}(\bb^0\to\rho^0\gamma)=\cases{0.9\times 10^{-6},  \cr 1.8
\times 10^{-6},  \cr}
\en
for $\eta=0.30$, $\rho=0.18$ (upper entry) and $\rho=-0.18$ (lower entry),
where we have applied Eq.(5) and the lifetimes $\tau(\bb^0)=1.50\times 10^
{-12}s$ and $\tau(B^-)=1.54\times 10^{-12}s$ [13]. It follows from (40) that
\be
R\equiv {\Gamma(B^-\to\rho^-\gamma)\over \Gamma(\bb^0\to\rho^0\gamma)}=\cases{
1.6,   \cr 2.3,   \cr}~~~{\rm for}~\rho=\cases{0.18,   \cr  -0.18,    \cr}
\en
and $\eta=0.30$. Hence, violation of isospin symmetry for $\bb\to\rho\gamma$
decay rates is at the level of 20\%. Since $R=2$ due to the electromagnetic
penguin contribution, any deviation of $R$ away from 2 gives the indicator
of long-distance effects.

\vskip 0.4cm
\centerline{\small Table I. The ratios of long- and short-distance
contributions}
\centerline{\small to the form factors $f_1$ and $f_2$ in
$\bb\to\rho\gamma$ decays for}
\centerline{\small  $\rho=0.18$ (first entry) and $\rho=-0.18$ (second
entry).}
\begin{center}
\begin{tabular}{|c||c|c|} \hline
 & $B^-\to\rho^-\gamma$ & $\bb^0\to\rho^0\gamma$ \\ \hline
$\left|f_1^{\rm VMD}/f_1^{\rm peng}\right|$ & 0.18~~0.15 & 0.20~~0.13 \\ \hline
$\left|f_1^{\rm pole}/f_1^{\rm peng}\right|$ & 0.21~~0.14 & 0.01~~0.01\\ \hline
$\left|f_2^{\rm VMD}/f_2^{\rm peng}\right|$ & 0.10~~0.09 & 0.11~~0.08 \\ \hline
$\left|f_2^{\rm pole}/f_2^{\rm peng}\right|$ & 0.16~~0.11 & 0.02~~0.02\\ \hline
\end{tabular}
\end{center}
\vskip 0.3 cm

 As for the $\bb^0\to D^{*0}\gamma$ decay, we find from (30) and (31) that
\be
{f_1^{\rm VMD}\over f_1^{\rm pole}}=1.33\,,~~~~~{f_2^{\rm VMD}\over f_2^{\rm
pole}}=0.09\,.
\en
We see that the form factor $f_2$ is dominated by the pole
contribution, while the VMD effect plays an essential role in $f_1$. This is
ascribed to the fact that, as can be seen from Eq.(30), there is a large
cancellation in $f_1^{\rm pole}$.
Since the decay rate is proportional to $|f_1|^2
+4|f_2|^2$ and $f_1^{\rm pole}/f_2^{\rm pole}=0.25$, it is easily seen that
the branching ratio of $\bb^0\to D^{*0}\gamma$
\be
{\cal B}(\bb^0\to D^{*0}\gamma)=\,0.93\times 10^{-6},
\en
is overwhelmingly dominated by the pole diagrams.
In the absence of the VMD contributions, this branching ratio will become
$0.74\times 10^{-6}$.
\footnote{This number is slightly different from the result $0.92\times
10^{-6}$ obtained in Ref.[1] since $a_2$ there is identified with ${1\over 2}
(c_--c_+)$.}

\vskip 0.4cm
\noindent{\bf 5. Discussion and Conclusion}
\vskip 0.25cm
   Assuming the validity of the VMD concept,
we have studied in the present paper the effect of VMD on the weak
radiative decays $\bb\to\rho\gamma$ and $\bb^0\to D^{*0}\gamma$.
Based on the factorization approach, we found
that $\bb\to\rho\gamma$ is dominated by the short-distance penguin
diagram and that the VMD contribution is ${\cal O}(10-20\%)$ of the
penguin amplitude. However, contrary to $B\to K^*\gamma$, the long-range
pole effect in $B^-\to\rho^-\gamma$ decay is comparable to the VMD one. The
pole contribution in $B\to K^*\gamma$ is suppressed due to the smallness of
the weak mixing $V_{ub}V_{us}$ relative to $V_{cb}V_{ud}$ appearing in the VMD
process. In the decay $B^-\to\rho^-\gamma$, the mixing matrix elements
entering into the pole diagram and the VMD diagram with $\rho^0$ and $\omega$
intermediate states are the same. However, the pole effect in $\bb^0\to
\rho^0\gamma$ is suppressed again (see Table I) owing to the fact that
the $W$-exchange diagram is color suppressed. Therefore,
as far as the relative magnitudes of the short- and long-distance
contributions are concerned, $\bb^0\to\rho^0\gamma$ resembles most to
$B\to K^*\gamma$.

    The branching ratio of $\bb\to\rho\gamma$, estimated to be of order
$10^{-6}$, depends strongly on the sign of the Wolfenstein parameter $\rho$. A
measurement of the ratio $R\equiv\Gamma(B^-\to\rho^-\gamma)/\Gamma
(\bb^0\to\rho^0\gamma)$ is of great interest for this purpose. Since the
short-distance penguin effect alone yields $R=2$, any deviation of $R$
from 2 will provide information
on the long-distance contribution and the sign of $\rho$.
We found that $R>2$ for $\rho<0$ and $R<2$ for $\rho>0$.

    The decay $\bb^0\to D^{*0}\gamma$ receives only long-distance
contributions. It turns out that though the VMD and pole diagrams contribute
comparably to the parity-conserving amplitude of $\bb^0\to D^{*0}\gamma$,
the parity-violating part is largely dominated by $W$-exchange bremsstrahlung.
Consequently, its branching ratio, predicted to be $0.9\times 10^{-6}$,
is overwhelmingly dominated by the pole contributions.

\vskip 2.0cm
\centerline{\bf ACKNOWLEDGMENTS}
\vskip 0.5cm
   I wish to thank Dr. D. L. Ting for drawing the graphs.
    This work was supported in part by the National Science Council of ROC
under Contract No. NSC84-2112-M-001-014.

\pagebreak
\centerline{\bf REFERENCES}
\vskip 0.3 cm
\begin{enumerate}

\item H.Y. Cheng, C.Y. Cheung, G.L. Lin, Y.C. Lin, T.M. Yan, and H.L. Yu,
CLNS 94/1278, hep-ph/9407303, to appear in Phys. Rev. D (1995).

\item E. Golowich and S. Pakvasa, \pl {\bf B205}, 393 (1988).

\item N.G. Deshpande, J. Trampetic, and K. Panose, \pl {\bf B214}, 467 (1988).

\item E. Golowich and S. Pakvasa, UMHEP-411, UH-511-800-94, hep-ph/9408370
(1994).

\item M.R. Ahmady, D. Liu, and Z. Tao, IC/93/26, hep-ph/9302209 (1993).

\item J.M. Soares, \pr {\bf D49}, 283 (1994); S. Narsion, \pl {\bf B327}, 354
(1994); A. Ali, V. Braun, and H. Simma, CERN-TH 7118/93, hep-ph/9401277 (1993).

\item D. Atwood, B. Blok, and A. Soni, SLAC-PUB-6635, hep-ph/9408373 (1994).

\item M. Misiak, \pl {\bf B269}, 161 (1991); \np {\bf B393}, 23 (1993); G.
Cella, G. Curci, G. Ricciardi, and A. Vicer\'e, \pl {\bf B248}, 181 (1990);
{\sl ibid.} {\bf B325}, 227 (1994); B. Grinstein, R. Springer, and M.B. Wise,
\np {\bf B339}, 269 (1990); R. Grigjanis, P.J. O'Donnell, M.
Sutherland, and H. Navelet, \pl {\bf B213}, 355 (1988); {\sl ibid.}
{\bf B286}, 413(E) (1992); {\sl Phys. Rep.} {\bf 228}, 93 (1993).

\item N. Isgur and M. Wise, \pr {\bf D42}, 2388 (1990).

\item M. Wirbel, B. Stech, and M. Bauer, \zp {\bf C29}, 637 (1985).

\item P.J. O'Donnell and H.K.K. Tung, \pr {\bf D48}, 2145 (1993).

\item M. Bauer, B. Stech, and M. Wirbel, \zp {\bf C34}, 103 (1987).

\item Particle Data Group, \pr {\bf D50}, 1173 (1994).

\item CLEO Collaboration, M.S. Alam {\it et al.,} \pr {\bf D50}, 43 (1994).

\item H.Y. Cheng and B. Tseng, IP-ASTP-21-94, hep-ph/9409408 (1994).

\item CDF Collaboration, FERMILAB Conf-94/127-E (1994).

\item CLEO Collaboration, R. Ammar {\it et al.,} \prl {\bf 71}, 674 (1993).

\item M. Neubert, \pl {\bf B338}, 84 (1994).

\item L. Wolfenstein, \prl {\bf 51}, 1945 (1983).

\end{enumerate}

\pagebreak
\centerline{\bf Figure Captions}
\vskip 1.0 cm
\begin{enumerate}

\item VMD processes contributing to $B^-\to\rho^-\gamma$ with the vector-meson
intermediate states $\j,~\psi',~\rho^0$ and $\omega$.

\item Same as Fig.1 except for $\bb^0\to\rho^0\gamma$.

\item $W$-annihilation diagram contributing to $B^-\to\rho^-\gamma$ and
$W$-exchange to $\bb^0\to\rho^0\gamma$. Contributions due to
photon emission from other quarks are denoted by ellipses.

\end{enumerate}

\end{document}